# Finding a Maximum Clique using Ant Colony Optimization and Particle Swarm Optimization in Social Networks

Mohammad Soleimani-Pouri
Department of Electrical, Computer and Biomedical Engineering, Qazvin branch, Islamic Azad University, Qazvin, Iran
m.soleimani@qiau.ac.ir

Alireza Rezvanian
Computer and Information Technology Engineering Department
Amirkabir University of Technology, Tehran, Iran
a.rezvanian@aut.ac.ir

Mohammad Reza Meybodi
Computer and Information Technology Engineering Department
Amirkabir University of Technology, Tehran, Iran
mmeybodi@aut.ac.ir

*Abstract*—Interaction between users in online social networks plays a key role in social network analysis. One on important types of social group is full connected relation between some users, which known as clique structure. Therefore finding a maximum clique is essential for some analysis. In this paper, we proposed a new method using ant colony optimization algorithm and particle swarm optimization algorithm. In the proposed method, in order to attain better results, it is improved process of pheromone update by particle swarm optimization. Simulation results on popular standard social network benchmarks in comparison standard ant colony optimization algorithm are shown a relative enhancement of proposed algorithm.

## I. Introduction

Today online social networks are formed a new type of life due to some facilities and services for a wide ranges of ages such as young to old. There is no doubt about either influence or growth of social networks. Therefore, several many of researchers are focused on social network analysis aspects. It seems to be useful, studying structure of relation between users in social networks. One of the important user group structure associated with a full connected of some users, which known as clique structure [1]. Several applications of finding clique are reported by researchers such as social networks analysis [2], online shopping recommendation [3] and evolution of social networks [12]. In literature, finding clique is categorized as NP-complete problems in graph theory [4].

Evolutionary algorithms (EAs) are stochastic optimization techniques based on the principles of natural evolution which applied several NP problems [13-15] such as clique problem. Various types of algorithms have been presented to solve clique problem, while evolutionary algorithms such as genetic algorithm (GA) and ant colony optimization (ACO) have been used more. Popular algorithm named Ant-clique algorithm, which make a maximal clique using sequential greedy heuristics based on ant colony optimization by adding vertices to the partial cliques iteratively [5]. Besides, another ACO based method hybridized by simulated annealing (SA) [6] and tabu search [7]. Although new hybrid algorithm obtained good results, they have a high complexity in practice.

In this study, Particle Swarm Optimization (PSO) algorithm has been applied as the heuristic to enhance the performance of ACO algorithm for finding a maximal clique in social network graph. Simulation results on social network benchmark are shown the better results in comparison with standard ACO algorithm. In the rest of this paper, section II and III are consisted of ACO and PSO introduction respectively, in section IV, proposed method is discussed. Simulation results on social networks datasets are reported in section V.

## II. Ant Colony optimization (ACO)

Ant Colony optimization (ACO) algorithm works well for solving several discrete problems. The basic algorithm of ACO was proposed by Dorigo as a multi agent approach in order to solve traveling salesman problem (TSP) [8]. The main idea of ACO algorithm is inspired from the behavior of seeking out food by colonies of ants. Ants search their environment randomly to find food. They return some of the food to their nest once found a food and leave pheromone in their return path. The amount of pheromone left on their path depends on quality and size of the food source and it evaporates gradually. Remaining pheromones will persuade other ants to follow the path and just after a short time, majority of the ants will trace the shorter path which is marked with stronger pheromone. Procedure of ACO algorithm has been presented in Figure 1.

```
Procedure ACO_MetaHeuristic
   while(termination_conditions)
      generateSolutions()
      daemonActions() {Optional}
      pheromoneUpdate()
   end while
end procedure
```

Figure 1. Pseudo-code of ACO algorithm [9]

During running of the algorithm, ants first produce different solutions randomly in the main loop after initialization. Afterwards, the solutions are improved by updating the pheromones and using a local search optionally. According to the problem and graph traverse, pheromones set on vertices or edges of graph. Traversing the edge between vertices i and j depends on the probability of edge which is calculated as below:

$$p_{ij}^k = \frac{(\tau_{ij}^\alpha)}{\sum(\tau_{ij}^\alpha)} \quad (1)$$

Where, $p_{ij}^k$ is probability of traversing the edge between vertices $i$ and $j$, while $\tau_{ij}^\alpha$ is amount of pheromone present on the above mentioned edge. An optional local search can contribute to improvement of the results prior to updating the pheromones. However, method of updating the pheromones can be like this:

$$\tau_{ij}^{t+1} = (1-\rho)\tau_{ij}^t + \Delta\tau_{ij}^t \quad (2)$$

Where, $\rho$ is evaporation rate of pheromone, $\tau_{ij}^{t+1}$ is amount of new pheromone for edge between $i$ and $j$, $\tau_{ij}^t$ is amount of current pheromone for edge between $i$ and $j$, $\Delta\tau_{ij}^t$ is amount of reinforced pheromone for proper solutions which can be calculated from the following equation.

$$\Delta\tau_{ij}^t = \begin{cases} 1 & if \quad \tau_{ij}^t \in good\ solution \\ 0 & Otherwise \end{cases} \quad (3)$$

### III. PARTICLE SWARM OPTIMIZATION (PSO)

Particle swarm optimization (PSO) is a population based optimization technique developed by Eberhart and Kennedy in 1995, which inspired from social behavior of birds seeking for food. In this method, group of birds seek for food in a specified space randomly. Birds follow the bird with the shortest distance to food in order to find position of the food [10].

Every solution in PSO algorithm which is called a particle corresponds to a bird in their pattern of social movement. Each particle has a value of fitness calculated by a fitness function. A particle bears greater fitness once it is located in a closer position to the target (food in the model of moving birds) within search space. Moreover, any particle represents a speed which is responsible to direct the particle. A particle will keep going through the problem space by following the optimum particle at current state [16].

A group of particles (solutions) are created randomly at the beginning of particle swarm optimization algorithm and they try to find the optimum solution through being updated among generations. A particle is updated in each step using the best local and global solution. The best position a particle has ever succeeded to reach is called *pbest* and saved while the best position achieved by the population of particles is named *gbest*. Velocity and location of each particle will be updated using Equations (4) and (5) in each step of implementing the algorithm after finding the best local and global values.

$$v_i^{t+1} = wv_i^t + c_1r_1\left(pbest_i^t - x_i^t\right) + c_2r_2\left(gbest_i^t - x_i^t\right) \quad (4)$$

$$x_i^{t+1} = x_i^t + v_i^t \quad i = 1,...,m \quad (5)$$

Where, $v_i$ is the velocity of $i^{th}$ particle and $x_i$ is the current position of it. $r_1$ and $r_2$ are random values in the range of (0,1). $c_1$ and $c_2$ are learning parameters usually assumed equal ($c_1=c_2$). $w$ is inertia weight which is considered as a constant or variable coefficient as random, linear, nonlinear and adaptive [11]. PSO has been used in various applications and this research utilizes it to improve amount of pheromones.

### IV. PROPOSED ALGORITHM

High complexity was a major drawback of the previous heuristic methods for solving the clique problem since it significantly adds to the volume of calculations. All methods provided so far apply the following relation to calculate $\Delta\tau$ although the proposed algorithm take the advantage of PSO algorithm to improve results and reduce complexity.

$$\Delta\tau^{t+1} = \frac{1}{(1+|G\text{-best}|-|\text{best-tour}|)} \quad (6)$$

This algorithm has used the hybrid of ACO and PSO algorithms in order to find the maximal clique in a graph. For this purpose, some ants are placed initially on the graph and follow paths to find the maximal clique. After evaporation of existing pheromones, proper path is updated by its amount on the edges using particle swarm optimization algorithm. This procedure is repeated until the optimum clique is obtained on the desired graph. Determining the amount of pheromone through PSO is such that the total pheromone measured at this step and the total pheromone associated with the best answer up to present step will be calculated taking into account the amount of pheromones at current step. Now, $\Delta\tau$ for the reinforced pheromone of the desired clique will be calculated with PSO algorithm using the following equation:

$$\Delta\tau^{t+1} = \Delta\tau^t + V^t \quad (7)$$

Where, $\Delta\tau_{ij}^t$ is amount of reinforcement for current pheromone and $\Delta\tau_{ij}^{t+1}$ is amount of reinforcement for new pheromone, while $V^t$ gives the amount of change which can be achieved from this equation:

$$V^{t+1} = c_1r_1(p\tau - \Delta\tau) + c_2r_2(g\tau - \Delta\tau) + c_3V^t \quad (8)$$

Where, $V^{t+1}$ is the new value of $V^t$. $r_1$ and $r_2$ are two random values within range of (0,1), while $c_1$, $c_2$ and $c_3$ are learning parameters ($c_1=c_2=c_3$). $p\tau$ and $g\tau$ are considered as the pheromone correspondent with the best current clique and the best clique found so far, respectively. In this case, discovering the changed amount of pheromone will be implemented much more intelligent.

Taking into consideration the mentioned items, general routine of proposed the algorithm can be summarized as below:
1. Initialize the parameters
2. Repeat steps 3 to 5 until reaching conditions of termination
3. Traverse the graph by ants and create sub-graphs as clique
4. Evaluate the obtained cliques
5. Update pheromones by PSO according to equation (8)
6. End.

## V. SIMULATION RESULTS

For evaluation of the proposed method, experiments applied on some popular social network datasets. The description of popular social networks is listed in the table 1.

TABLE I. SOCIAL NETWORKS DESCRIPTIONS

|      | Dataset | Nodes | Edges |
| --- | --- | --- | --- |
| I | Zachary's karate club | 34 | 78 |
| II | Common adjective and nouns in "David Copperfield" | 112 | 425 |
| III | Neural network of the nematode C. Elegans | 297 | 8479 |
| IV | social network of dolphins, Doubtful Sound, New Zealand | 62 | 159 |
| V | Pajek network: Erdos collaboration network 971 | 472 | 1314 |
| VI | Pajek network: Erdos collaboration network 991 | 492 | 1417 |
| VII | Pajek network: World Soccer, Paris 1998 | 35 | 295 |
| VIII | Pajek network: graph and digraph glossary | 72 | 118 |
| IX | Pajek network: Slovenian journals 1999-2000 | 124 | 823168 |
| X | co-authoship of scientists in network theory & experiments | 1589 | 1190 |
| XI | Pajek network: SmaGri citation network | 1059 | 4919 |
| XII | email interchange network, Univ. of Rovira i Virgili, Tarragona | 1133 | 5451 |

Topology of zachary's karate club and social network of dolphins are presented in figure 2 and 3 respectively as the most popular of the social network datasets.

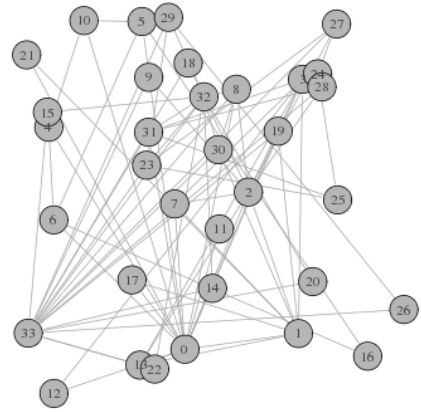

Figure 2. Zachary's karate club social network

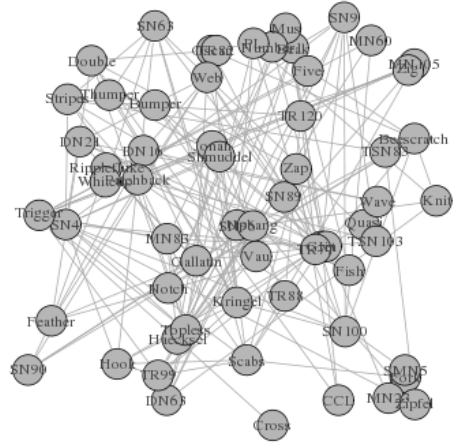

Figure 3. Dolphins social network

The setting of parameters for experiment is listed in below. It is noted that choosing different values for improving the results is also possible.

In this paper has used 30 ants with $\rho=0.95$, $\Phi=0.0002$, $\Delta\tau_{initial}=0$, $\tau_{min}=0.01$, $\tau_{max}=6$.

Meanwhile, parameters of PSO were initialized as $V=0$, $c_1=c_3=0.3$, and $c_2=1-c_1$. $\alpha$ and $\rho$ were given the following values based on the experiments done and $t$ is the number of iterations.

$$\alpha(t) = \begin{cases} 1 & t \leq 100 \\ 2 & 100 < t \leq 400 \\ 3 & 400 < t \leq 800 \\ 4 & t > 800 \end{cases} \quad (9)$$

$$\rho(t+1) = \begin{cases} (1-\varphi)\rho(t) \\ 0.95 & \text{if } \rho(t) > 0.95 \end{cases} \quad (10)$$

Average of 10 independent runs with 1000 iterations for each implementation have been listed in table 2 and table 3 for proposed method (ACO-PSO) and ACO algorithm respectively, including the maximum (Best), average (Avg), standard deviation (Std) and run-time of algorithm for finding a maximum clique in each graph datasets.

TABLE II. SIMULATION RESULTS OF ACO-PSO FOR FINDING A MAXIMUM CLIQUE

| Graph | Best | Avg | Std | Run-time |
|---|---|---|---|---|
| I | 5 | 4.995 | 0.070 | 11.84 |
| II | 5 | 4.783 | 0.412 | 23.37 |
| III | 7 | 5.543 | 1.482 | 49.47 |
| IV | 5 | 4.998 | 0.044 | 15.69 |
| V | 7 | 5.848 | 0.6673 | 105.11 |
| VI | 7 | 6.011 | 0.819 | 111.89 |
| VII | 5 | 4.118 | 0.322 | 10.742 |
| VIII | 4 | 3.985 | 0.121 | 12.94 |
| IX | 4 | 3.185 | 0.299 | 22.07 |
| X | 3 | 2.253 | 0.371 | 118.59 |
| XI | 8 | 6.409 | 0.765 | 481.34 |
| XII | 12 | 7.997 | 2.241 | 529.04 |

TABLE III. SIMULATION RESULTS OF ACO FOR FINDING A MAXIMUM CLIQUE

| Graph | Best | Avg | Std | Run-time |
|---|---|---|---|---|
| I | 5 | 4.991 | 0.082 | 64.03 |
| II | 5 | 4.709 | 0.495 | 355.81 |
| III | 7 | 5.521 | 1.568 | 1121.71 |
| IV | 5 | 4.991 | 0.094 | 132.12 |
| V | 7 | 8.753 | 0.861 | 5853.59 |
| VI | 7 | 5.961 | 0.924 | 6281.35 |
| VII | 5 | 3.910 | 0.293 | 55.15 |
| VIII | 4 | 3.943 | 0.232 | 153.66 |
| IX | 4 | 3.017 | 0.316 | 438.87 |
| X | 3 | 2.175 | 0.434 | 2142.03 |
| XI | 8 | 6.025 | 0.815 | 9153.65 |
| XII | 12 | 7.562 | 2.374 | 11109.17 |

Table 2 and table 3 indicate that the proposed method (ACO-PSO) produces better results in comparison with ACO method since the proposed approach is an appropriate method to update pheromones of the traversed paths for ants in calculating the optimum clique.

## VI. CONCLUSION

A new hybrid algorithm has been presented in this paper using ACO and PSO (ACO-PSO) for finding a maximum clique in social networks. Traditional algorithms suffered high complexity while the hybrid proposed algorithm just change the process of update pheromone. It has been shown that the new algorithm was able to improve the basic ACO algorithm as simply and quickly. Simulation results on popular social networks datasets indicate the improved results for proposed algorithm in comparison with the ACO algorithm.